\documentclass[%
 preprint,
 amsmath,amssymb,
 aps, physrev,
]{revtex4-2}

\usepackage{graphicx}
\usepackage{dcolumn}
\usepackage{bm}
\usepackage{comment}
\usepackage{float}
\usepackage{subcaption}

\usepackage{tikz}
\usetikzlibrary{calc}

\begin{document}

\preprint{}

\title{\textbf{Deterministic single-electron trapping on solid neon using engineered dielectric surface geometry.} 
}%

\author{Kundan Surse}
    \email{Corresponding author: k.surse@unsw.edu.au}
\author{Eric Helgemo}
\author{Andrew Palmer}
\author{Md Serajum Monir}
\author{Lukas Delventhal}
\author{Thanh Nguyen}
\author{Maja Cassidy}
\author{Rajib Rahman}
    \email{Corresponding author: rajib.rahman@unsw.edu.au}

\affiliation{School of Physics, University of New South Wales, Sydney, NSW 2052, Australia}


\date{\today}

\begin{abstract}
Levitating electron qubit on the surface of solid neon has recently emerged as a promising and intrinsically noise-resilient platform for quantum information processing. Their ultra-clean, inert environment suppresses conventional decoherence pathways associated with lattice disorder, charge traps, and nuclear-spin baths that limit coherence in semiconductor qubits. Yet, uncontrolled surface features such as bumps, valleys, and electrode-defined gaps can bind electrons unintentionally, contributing charge noise and inducing spin–orbit-coupling-mediated decoherence. To address this challenge, we propose an engineered interface in which a dielectric layer is deposited beneath the solid neon to provide an atomically smooth template, eliminating surface-roughness-induced trapping. By selectively etching this dielectric layer at desired qubit locations, deterministic potential minima can be engineered to reliably capture electrons while suppressing unwanted surface bound states. We perform large-scale Schrodinger and Poisson simulation to compare the existing and proposed strategies of electron trapping on neon, obtaining good agreement with recent experimental measurements.   
\end{abstract}

\maketitle


\section{\label{sec:level1}Introduction}

Achieving both long coherence and fast operation remains a central challenge in quantum computing hardware. Existing platforms often realize only one of these advantages strongly: trapped ions offer excellent coherence and gate fidelities but are relatively slow and difficult to scale, while solid-state platforms provide compact architectures, fast electrical control, and compatibility with nanofabrication, but suffer from charge noise, dielectric loss, interface disorder, strain, and complex semiconductor--oxide--metal environments \cite{Ezratty2023,Wang2023,Flamini2018,Bruzewicz2019,Browaeys2020,Veldhorst2023,LossDivincenzo1998,kane1998,Zwanenburg2013}. This has motivated hybrid approaches that combine the cleanliness of atomic systems with the speed and scalability of on-chip electronics. Electrons levitating above cryogenic dielectric surfaces, especially liquid helium and solid neon, provide such a setting because the active quantum degree of freedom resides primarily in vacuum rather than inside a disordered material host \cite{Platzman1967,Lyon2006}. This suppresses many conventional decoherence channels while still enabling electrical confinement, microwave control, and integration with superconducting circuit-QED architectures \cite{Zhou2022,Zhou2023,Zhou2024}.

Solid neon is particularly attractive because of its material simplicity. Neon is a closed-shell noble element, with electronic configuration \(1s^2\,2s^2\,2p^6\), that crystallizes through weak van der Waals interactions into an ultraclean insulating solid with a face-centered-cubic lattice and a lattice constant of roughly \(4.4\text{--}4.6~\mathrm{\AA}\) \cite{Drummond2008,Endoh1975,Batchelder1967}. Its large band gap, of order \(21~\mathrm{eV}\), implies negligible electrical conduction, while its small relative permittivity, \(\varepsilon_r \approx 1.245\), makes it only a weak perturbation from vacuum \cite{Boursey1970,Bernstorff1986,Saile1979,Endoh1975}. Natural neon is also more than \(99.7\%\) nuclear-spin-free, since its dominant isotopes have zero nuclear spin. These properties make solid neon an exceptionally quiet dielectric medium compared with semiconductor quantum dots, where lattice disorder, charge traps, and nuclear-spin baths can limit qubit performance \cite{Sun2010,Pla2018,Connors2019,HuangHu2013,Cywinski2009}. In this platform, an electron is bound above the neon surface by image-charge attraction and short-range Pauli repulsion, while lateral electrostatic confinement from patterned electrodes defines a quantum dot above the film \cite{Kajita1984,Monarkha1975,Chen2022}.

Recent experiments have demonstrated rapid progress in solid-neon electron qubits. Single electrons have been trapped on solid neon and coupled to superconducting resonators, enabling strong charge--photon coupling, coherent manipulation, and dispersive readout \cite{Zhou2022}. More recent work has shown highly coherent charge qubits with coherence times approaching \(92.9~\mu\mathrm{s}\), a remarkable result for a charge-based qubit and evidence that the solid-neon environment can strongly suppress decoherence \cite{Zhou2022,Zhou2023,Zhou2024}. Beyond charge qubits, the long-term promise of this platform lies in spin qubits, where theoretical studies suggest that the clean neon environment can suppress key spin-dephasing mechanisms \cite{Chen2022}. However, spin qubits on solid neon have not yet been experimentally realized, and scalable operation will require deterministic loading, reproducible trapping, precise qubit positioning, and long-range coupling between distant spins through superconducting resonators. Such spin--photon coupling generally requires a strong electric dipole moment combined with controlled spin--charge hybridization \cite{Samkharadze2018}.

Despite rapid progress, deterministic electron trapping remains a major unresolved challenge. Electrons are typically loaded from a tungsten filament and captured probabilistically in available potential minima \cite{Zhou2022}. In practice, the trapping landscape can be shaped not only by intended electrode potentials but also by resonator gaps, etched trenches, film nonuniformity, and surface roughness. Recent experiments demonstrated that the resonator-trench depth and
substrate surface properties strongly influence the formation of electron-on-neon
charge states and their coupling to microwave resonators, with shallow etched
features producing the strongest coupling \cite{Zheng2025Morphology}.  Together
with theoretical predictions that weak surface curvature can generate bound
electronic states \cite{Kanai2024}, these observations establish nanoscale
morphology as an active component of the electron-confinement landscape rather
than merely a fabrication imperfection. This issue was highlighted in a recent NbTiN-resonator-based experiment, where the electron was not localized at the intended position of maximum charge--photon coupling and its location had to be inferred from differential coupling to nearby electrodes \cite{wang2026charge}. Because the electron is bound only a short distance above the surface, nanoscale morphology can strongly influence the confinement potential, orbital spectrum, dipole distribution, and coupling strength \cite{Kajita1984}. Rather than treating surface roughness only as disorder, this work explores engineered solid-neon surface geometry as a quantum design knob. Using large-scale quantum electrostatic simulations, we investigate how structured neon surfaces can create deterministic trapping sites, suppress parasitic localization, enhance confinement, and reshape the electronic wavefunction. Our results show that surface engineering provides a practical pathway from accidental confinement toward reproducible, scalable levitating-electron qubits on solid neon.

\section{Methodology}

\begin{figure}[H]
    \centering
    \begin{subfigure}[b]{0.48\linewidth}
        \centering
        \includegraphics[width=\linewidth]{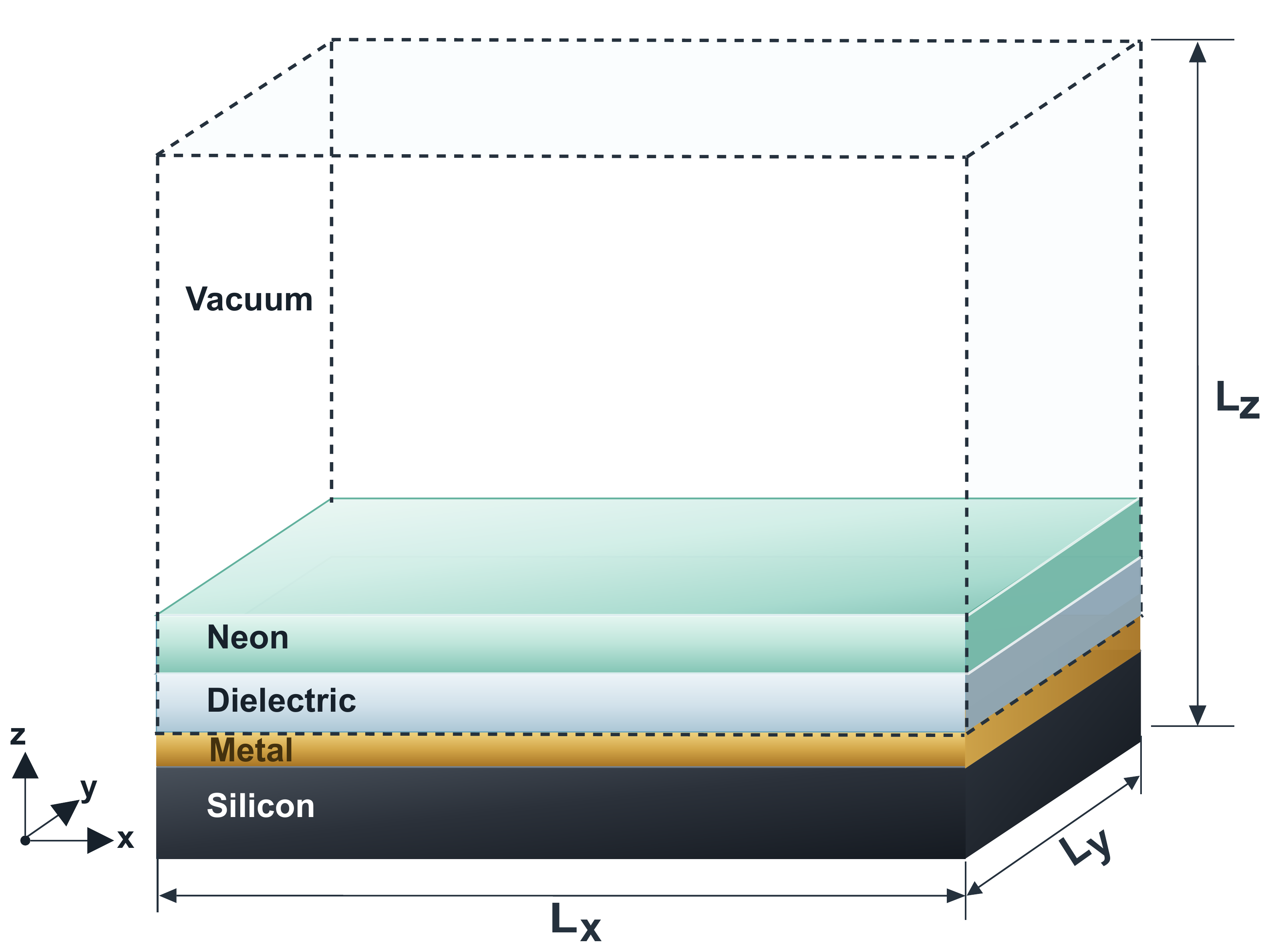}
        \caption{Simulation Domain}
        \label{fig:Sim_Domain}
    \end{subfigure}
    \hfill
    \begin{subfigure}[b]{0.48\linewidth}
        \centering
        \includegraphics[width=\linewidth]{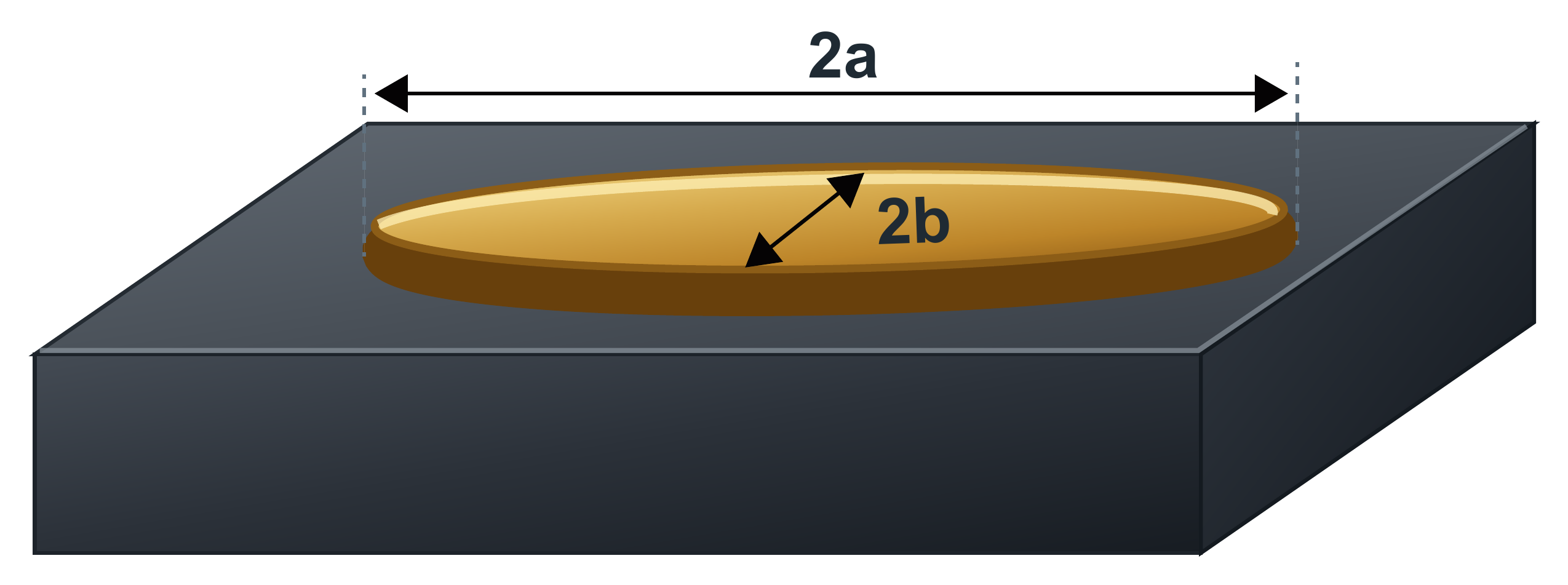}
        \caption{Device gate geometry.}
        \label{fig:device_gate_geometry}
    \end{subfigure}

    \caption{Device geometry used in the simulations: (a) Simulation domain shown in dotted box with material layers, and (b) patterned gate geometry showing an elliptical gate design with a and b as major and minor axes \cite{Zhou2022}.}
    \label{fig:device_geometry_subfigures}
\end{figure}

Motivated by the need to understand and control electron trapping on engineered solid-neon surfaces, this work uses numerical simulations to connect device geometry with the resulting electrostatic confinement and quantum states. We developed a three-dimensional finite-difference Schrödinger–Poisson framework for capturing the quantum electrostatics and wavefunctions of the system. The simulation domain was chosen large enough to capture the full electrostatic extent of the gate-defined trap while maintaining accurate spatial resolution near the dielectric interfaces, shown in Fig \ref{fig:Sim_Domain}. In the baseline implementation, the computational volume was
$ L_x \times L_y \times L_z = 2200~\mathrm{nm} \times 500~\mathrm{nm} \times 100~\mathrm{nm},$
discretized on a uniform mesh of $n_x \times n_y \times n_z = 121 \times 121 \times 111$.
This mesh was used consistently for electrostatics, material assignment, and quantum eigenstate calculation. The device coordinates were defined in such a way so that all dielectric interfaces, gate masks, and etched features could be encoded directly in real space. The quantum states of the levitating electron were computed by solving the single-particle Schr\"odinger equation

\begin{equation}
\left[
-\frac{\hbar^{2}}{2m_{e}} \nabla^{2}
- e\,V_q(\mathbf{r})
\right]
\Psi(\mathbf{r})
=
E\,\Psi(\mathbf{r}) .
\end{equation}
where \(m_e\) is the free electron mass and \(V_q(\mathbf{r})\) is the effective electron potential energy landscape. In the implementation, this quantum potential is written as
\begin{equation}
V_q(\mathbf{r}) = E_c(\mathbf{r}) - \phi(\mathbf{r}),
\end{equation}
where \(E_c(\mathbf{r})\) is the position-dependent conduction-band barrier determined by the material stack and \(\phi(\mathbf{r})\) is the electrostatic potential obtained from Poisson's equation ($E_c$ being zero in vaccum). The Laplacian operator was discretized using second-order finite differences along all three spatial directions, and the resulting sparse Hamiltonian was solved to obtain the lowest-energy bound states. In the Schrödinger–Poisson calculation, the electrostatic potential was first obtained from the device geometry and applied boundary conditions, and the corresponding quantum eigenstates were then computed. The ground-state wavefunction was used to evaluate the electron density for single-electron occupation, while the ground and excited orbital states were used to analyze level spacing, excited-state structure, and wavefunction localization.

Electrostatic potential was calculated by solving
\begin{equation}
 \nabla \cdot \left[\varepsilon(\mathbf{r}) \nabla \phi(\mathbf{r})\right] = -\rho(\mathbf{r})   
\end{equation}

with spatially varying permittivity \(\varepsilon(\mathbf{r}) = \varepsilon_0 \varepsilon_r(\mathbf{r})\) as defined in layer stack and device geometry. The Poisson operator was assembled in divergence form so that dielectric discontinuities across vacuum, neon, and dielectric layer were treated consistently. Dirichlet boundary conditions were imposed at the top, bottom, and side surfaces of the simulation box. The bottom boundary contained the patterned gate electrode, while the top boundary was fixed close to the bottom-gate bias to represent the far electrostatic environment, as the field drops over a large length but we are accounting for only 100~nm in z-direction. The side boundaries were grounded.

In the code, the bottom gate was modeled as an elliptical Dirichlet electrode centered at \((x_0 = L_x/2,y_0=L_y/2)\), with semi-axes $a_{\mathrm{gate}} = 1000~\mathrm{nm}$ and $b_{\mathrm{gate}} = 150~\mathrm{nm}$. The potential inside the ellipse at the lower boundary was set to $V_{\mathrm{bottom}} = 1.0~\mathrm{V}$, while the upper boundary was fixed at $V_{\mathrm{top}} = 0.99~\mathrm{V}$. This configuration creates a weak but extended vertical electric field while allowing the lateral confinement to be dominated by the patterned electrode. The material stack was defined in terms of dielectric permittivity and barrier potential. Vacuum was assigned \(\varepsilon_r=1.0\). Solid neon was assigned a relative permittivity $\varepsilon_r^{\mathrm{Ne}} = 1.2445$, and an effective barrier height $V_{\mathrm{Surf}}^{\mathrm{Ne}} = 0.7~\mathrm{eV}$ \cite{Zhou2022}. When dielectric spacer was included for single shot calculation of eigenfunction and eigen-energies it was treated using an effective scalar permittivity $\varepsilon_r^{\mathrm{dielectric}} = 3.5,$ with barrier $V_{\mathrm{Surf}}^{\mathrm{dielectric}} = 5.0~\mathrm{eV}$, otherwise it is considered as a sweep parameter.   A nominal dielectric thickness of $L_{\mathrm{dielectric}} = 10~\mathrm{nm}$ and a conformal neon coating thickness of $L_{\mathrm{Ne}} = 10~\mathrm{nm}$ were used in the reference geometry. The conformal nature of the neon overlayer was especially important: rather than depositing neon only as a vertical slab, the code implemented neon coverage over both top surfaces and sidewalls, which allowed etched or trenched structures to be coated realistically. This feature was essential for studying how the levitating electron responds to bumps, valleys, and etched confinement pockets.

The simulation workflow used an initial electrostatic guess obtained by linearly interpolating between the bottom-gate potential and the top boundary. Using this initial guess of potential, the Hamiltonian was constructed and the lowest quantum state was obtained. The probability density \(|\psi(\mathbf{r})|^2\) was normalized over the full three-dimensional volume. In the general implementation, this probability density can be converted into a charge density and fed back into Poisson's equation. However, in the present single-electron confinement study the solver was used primarily in the single-particle limit, and the final energy spectrum was extracted from the converged electrostatic landscape. The iterative loop used a convergence tolerance of \(10^{-4}\) with a maximum of 30 iterations. For the final eigenvalue solve, the lowest six states were extracted using a tighter eigensolver tolerance of \(10^{-10}\). From the converged states, we evaluated the qubit eigen-energies and splitting $E_i$ and $\Delta E = E_2 - E_1$ respectively. This $\Delta E$ was used throughout the study as a primary descriptor of confinement strength, orbital level spacing, and tunnelling-related coupling.

\section{Schrodinger-Poisson Model Validation}
\label{sec:validating_model}

The first stage of this study focused on validating the 3D Schr\"odinger--Poisson solver using the simplest experimentally relevant geometry: a smooth gate-defined trapping structure (Fig.~\ref{fig:device_gate_geometry}) comprising a solid neon layer above a buried electrode, without any additional dielectric layer. This baseline device isolates the electrostatic confinement generated by the gate and the vertical confinement imposed by the neon surface barrier, thereby providing a clean reference system for benchmarking the numerical model. The validation criteria were the low-lying eigenenergies, the spatial structure of the corresponding wavefunctions, the ground-state confinement widths, and the orbital splitting $E_2-E_1$, which defines the qubit transition frequency.

\begin{figure}[t]
    \centering

    \begin{subfigure}[b]{0.48\linewidth}
        \centering
        \includegraphics[width=\linewidth]{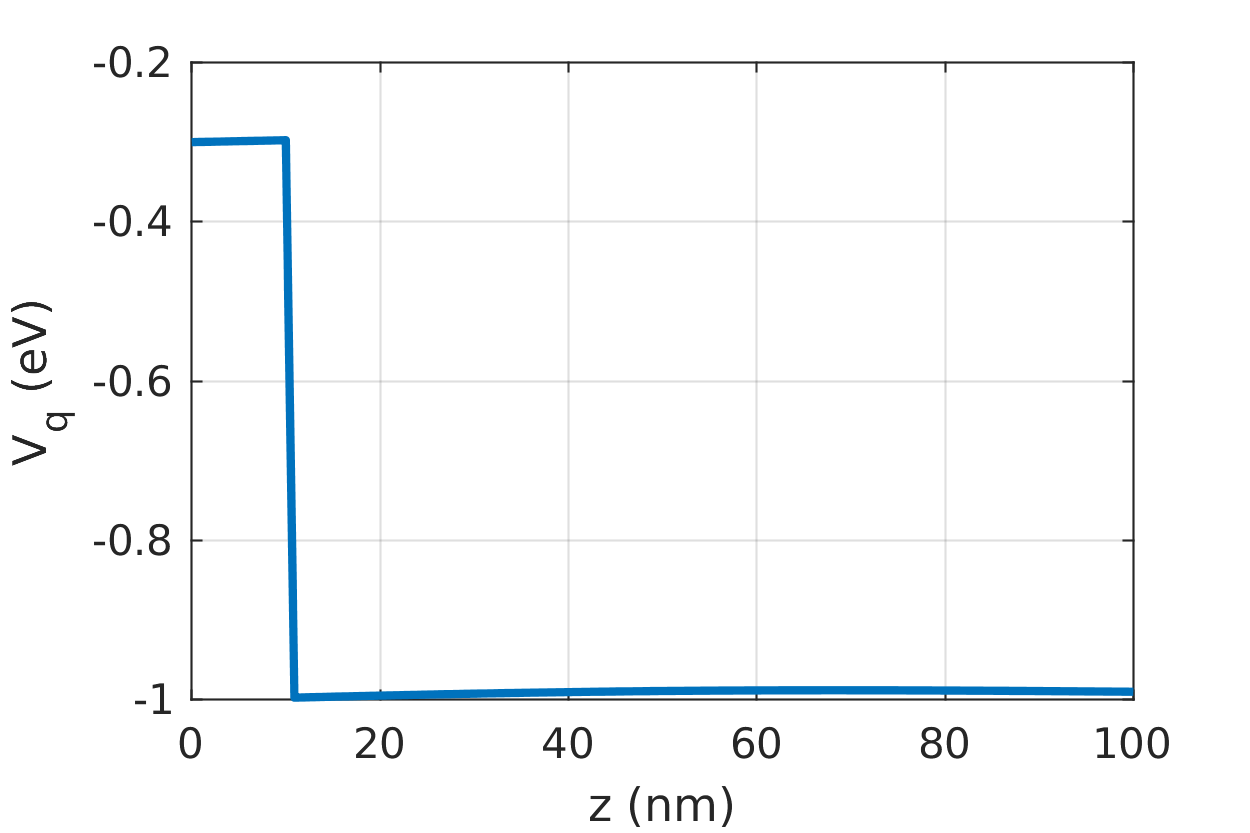}
        \caption{Potential along the central \(z\)-line.}
        \label{fig:pot1d_centerline}
    \end{subfigure}
    \hfill
    \begin{subfigure}[b]{0.48\linewidth}
        \centering
        \includegraphics[width=\linewidth]{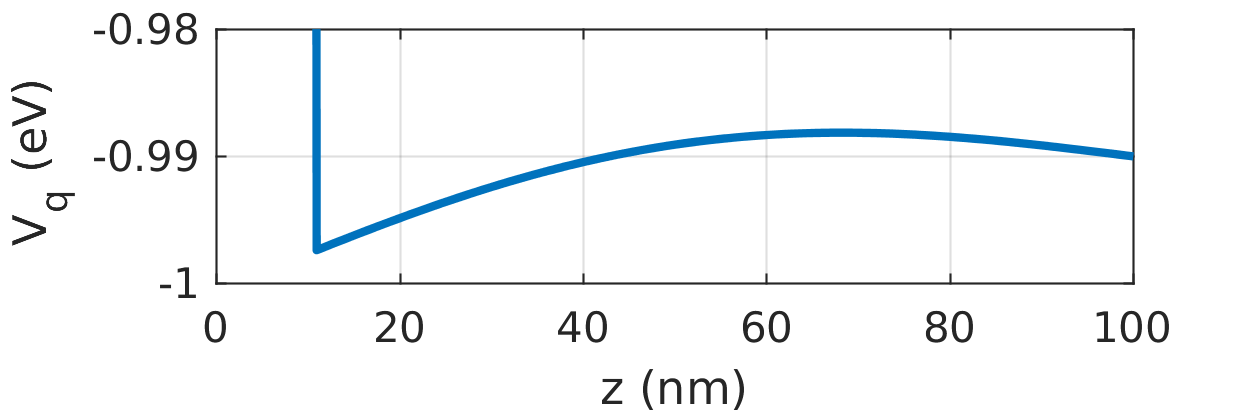}
        \caption{Zoomed view of the central \(z\)-line potential.}
        \label{fig:pot1d_centerline_zoom}
    \end{subfigure}
     \vspace{0.4cm}
     \begin{subfigure}[b]{0.48\linewidth}
        \centering
        \includegraphics[width=\linewidth]{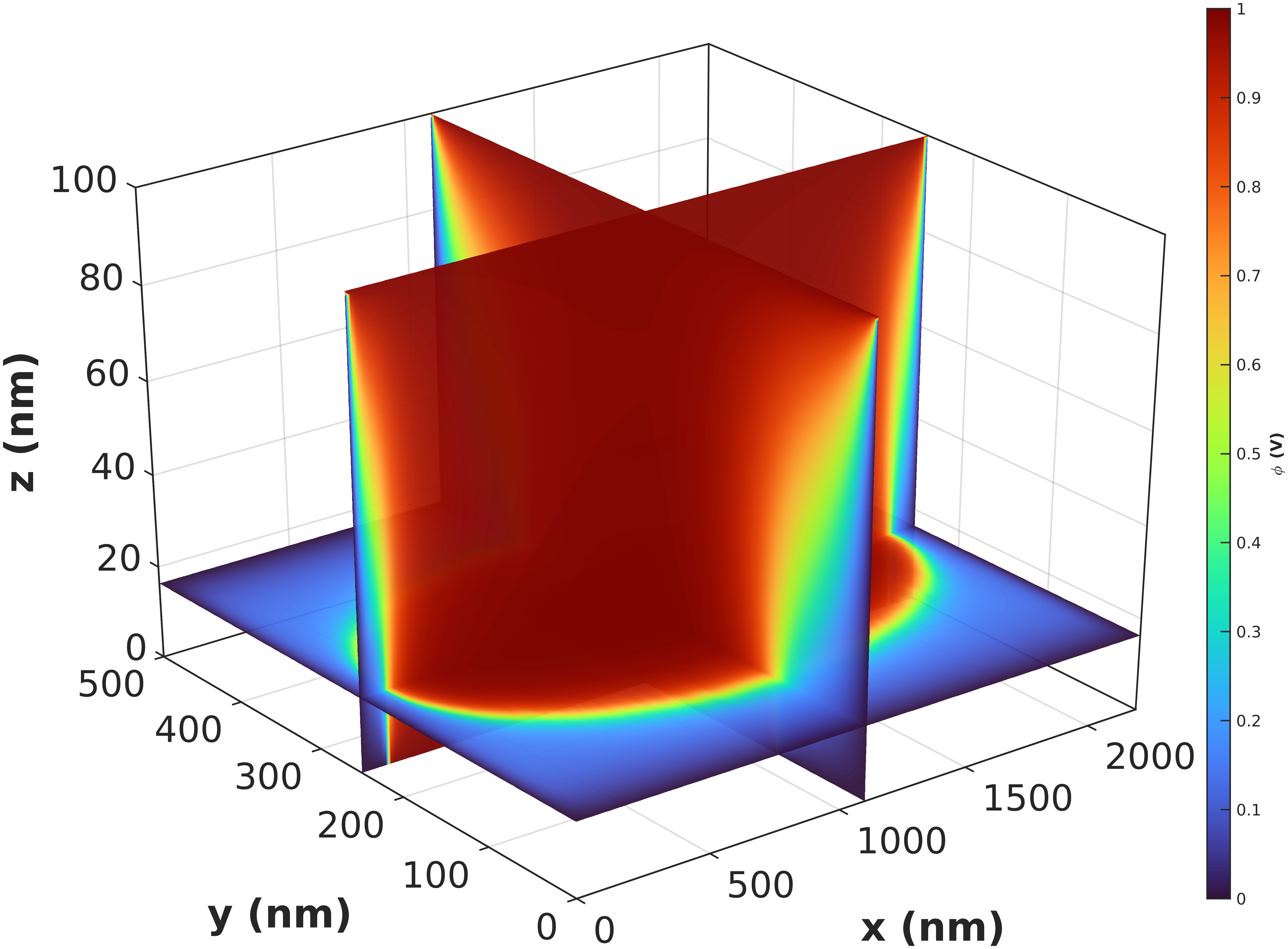}
        \caption{Orthogonal slices of the three-dimensional electrostatic potential.}
        \label{fig:pot3d_slices}
    \end{subfigure}
     \hfill
    \begin{subfigure}[b]{0.48\linewidth}
        \centering
        \includegraphics[width=\linewidth]{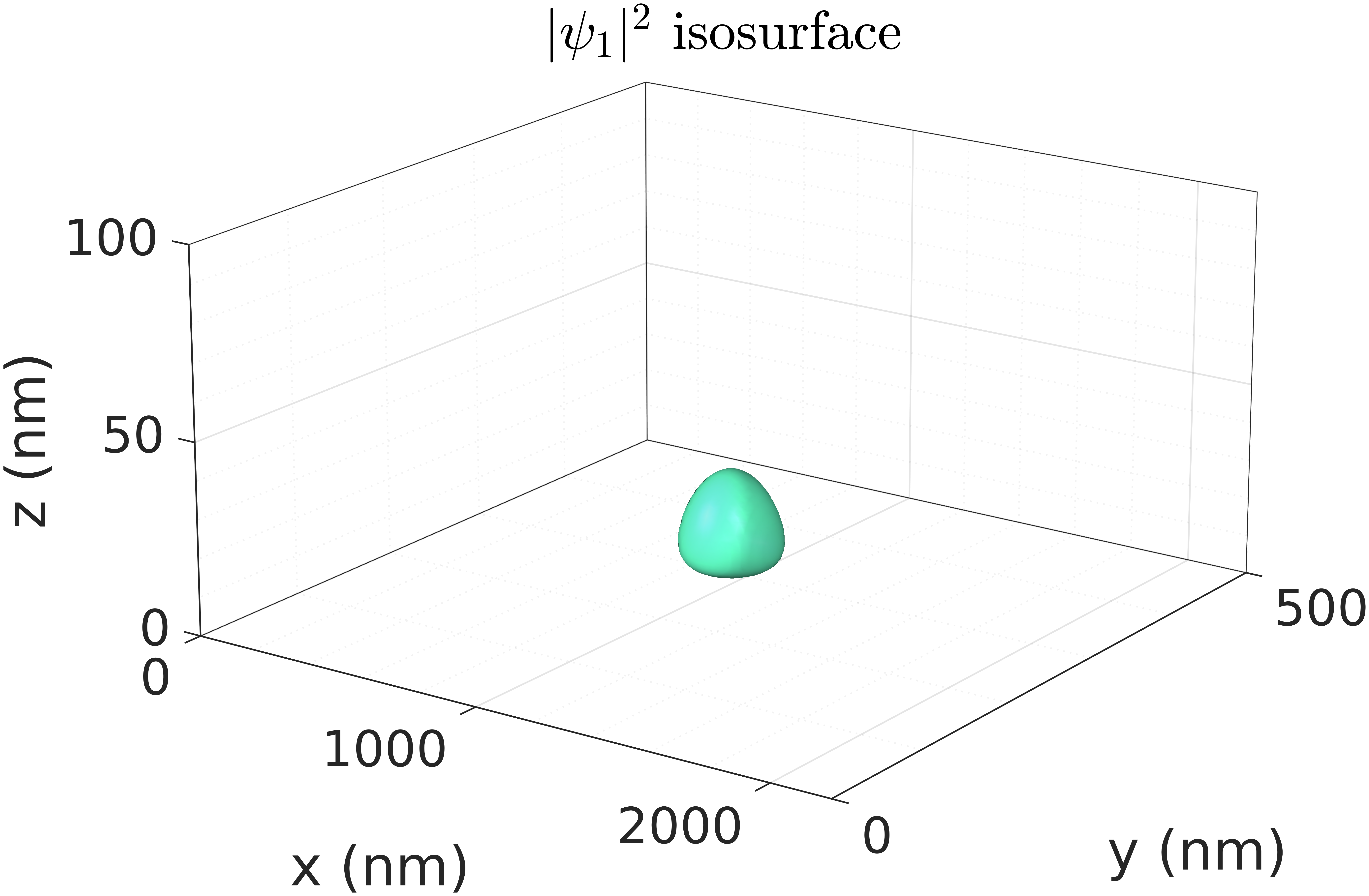}
        \caption{Three-dimensional ground-state wavefunction isosurface.}
        \label{fig:psi1_3d}
    \end{subfigure}

    \caption{Electrostatic potential landscape and ground-state wavefunction of the levitating-electron trap.}
    \label{fig:potential_wavefunction_combined}
\end{figure}

To ensure numerical convergence, simulations were performed on successively refined meshes of $41 \times 41 \times 111$, $101 \times 101 \times 111$, and $121 \times 121 \times 111$. The difference in the low-energy spectrum between the last two meshes was found to be below $1~\mu\mathrm{eV}$, and the $121 \times 121 \times 111$ grid was therefore adopted for all further calculations.

Fig.~\ref{fig:pot1d_centerline} and ~\ref{fig:pot1d_centerline_zoom} shows the one-dimensional electrostatic potential along the central $z$-line. The sharp step identifies the interfacial barrier, while the zoomed view reveals the shallow curvature of the trapping region responsible for vertical confinement. The full three-dimensional potential is shown in Fig.~\ref{fig:pot3d_slices}, where orthogonal slices demonstrate the formation of a laterally localized and vertically confined electrostatic well.

\begin{figure}[t]
    \centering
    \includegraphics[width=\linewidth]{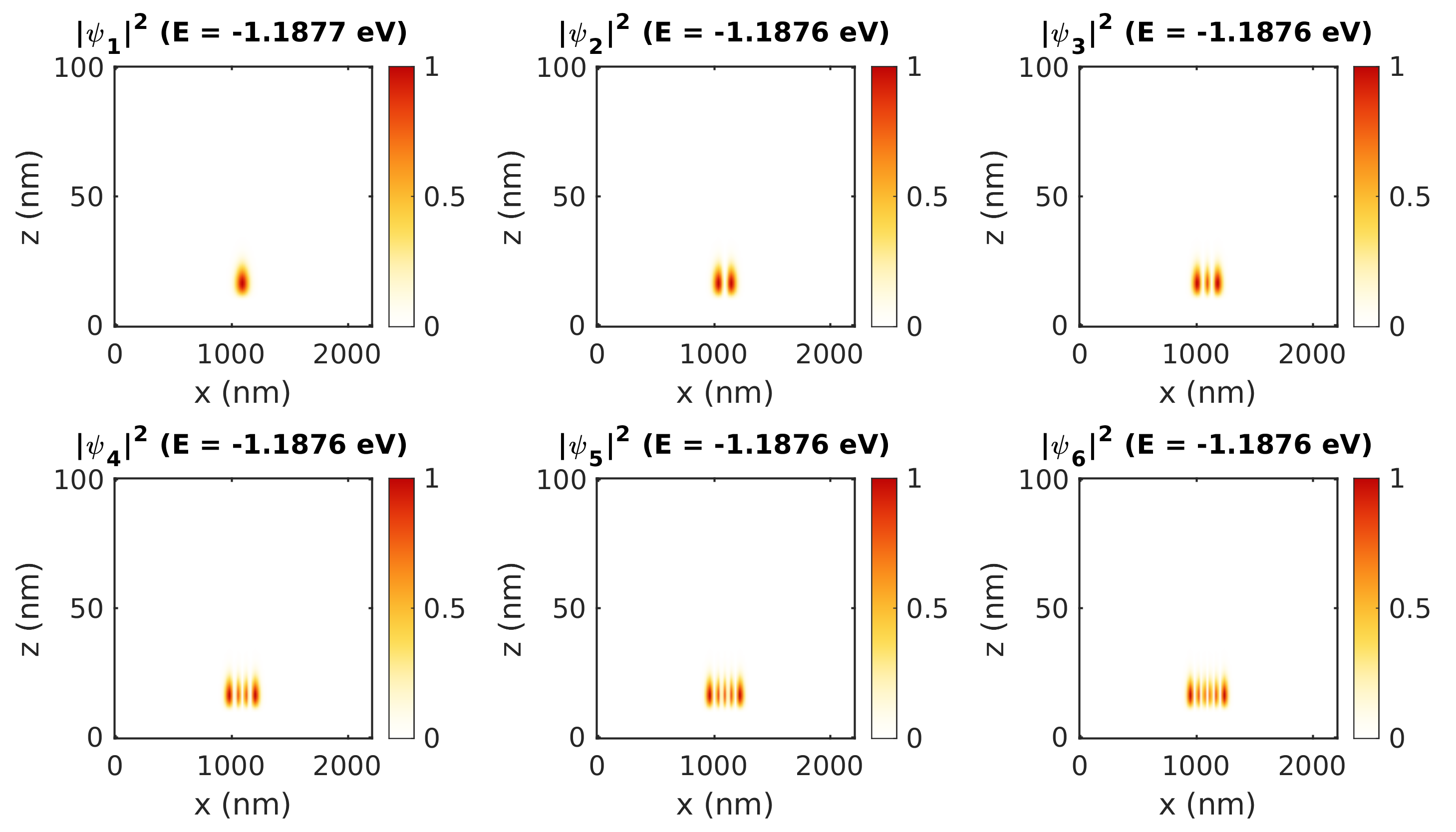}
    \caption{Probability-density distributions of the first six eigenstates, showing the progressive increase in nodal structure with excitation number.}
    \label{fig:wavefunctions_1to6}
\end{figure}

The corresponding 3D ground-state wavefunction is presented in Fig.~\ref{fig:psi1_3d} which confirms strong localization near the intended location centered on the gate. The first six eigenstates are shown in Fig.~\ref{fig:wavefunctions_1to6}. A systematic increase in nodal structure is observed primarily along the weakly confined in-plane direction, indicating that the low-lying spectrum is governed by anisotropic orbital excitations in a near-harmonic potential.

The corresponding eigenenergies and transition frequencies are listed in Table~\ref{tab:eigs6}. The fundamental orbital splitting is \(E_2-E_1 = 6.212\)~GHz, while the higher transition frequencies remain approximately evenly spaced, as expected for a smoothly confined single-electron trap. This value is in close agreement with previously reported eNe experimental splittings of \(\Delta E = 6.426\)~GHz~\cite{Zhou2022} and \(\Delta E = 6.3915\)~GHz~\cite{Zhou2024}, differing by only \(3.3\%\) and \(2.8\%\), respectively. Finally, Gaussian fits to the ground-state marginal densities, yield confinement widths of \(\sigma_x = 37.95\)~nm, \(\sigma_y = 8.70\)~nm, and \(\sigma_z = 3.66\)~nm. These values confirm a strongly anisotropic confinement profile, with particularly harmonic oscillator type trap potential.

Taken together, these results show that our numerical framework captures the essential low-energy physics of experimentally demonstrated gate-defined levitating-electron confinement on solid neon~\cite{Zhou2022,Zhou2024}. This agreement validates the model and establishes it as a reliable reference point for the more complex engineered structures examined in the following sections.

\begin{table}[H]
\centering
\caption{Lowest six eigenenergies and their corresponding transition frequencies from the ground state. The frequency is computed as $\Delta f_{n1}=(E_n-E_1)/h$, using $1~\mathrm{eV}/h = 2.417989242\times10^{5}~\mathrm{GHz}$.}
\label{tab:eigs6}
\begin{tabular}{c c c}
\hline
State $n$ & $E_n$ (eV) & $\Delta f_{n1}$ (GHz) \\
\hline
1 & $-1.187675292156$ & $ - $ \\
2 & $-1.187649601197$ & $6.212$ \\
3 & $-1.187626124539$ & $11.889$ \\
4 & $-1.187602605718$ & $17.576$ \\
5 & $-1.187578400973$ & $23.428$ \\
6 & $-1.187553538478$ & $29.440$ \\
\hline
\end{tabular}
\end{table}


\section{Surface Roughness Analysis }
After validating the baseline gate-defined device, controlled geometric perturbations were introduced to investigate how local surface topography modifies electron confinement. Positive and negative deformations of the neon surface, modeled as nanoscale bumps and valleys at the dielectric interface, were considered. This analysis is motivated by the work of Kanai \textit{et al.}~\cite{Kanai2024}, who showed that weak curvature on solid neon can generate bound electronic states through curvature-induced electrostatic potentials arising from the surface-charge distribution. In particular, surface protrusions can support annular or ``quantum-ring''-like states, while depressions and even atomic-scale step features may also act as unintended trapping centers. Such disorder-induced localization is highly relevant for qubit operation, since electrons may become pinned to random surface irregularities rather than the intended electrostatic trap, thereby introducing charge noise and additional decoherence pathways.
\begin{figure}[t]
    \centering
    \begin{subfigure}[t]{0.4\linewidth}
        \centering
        \includegraphics[width=\linewidth]{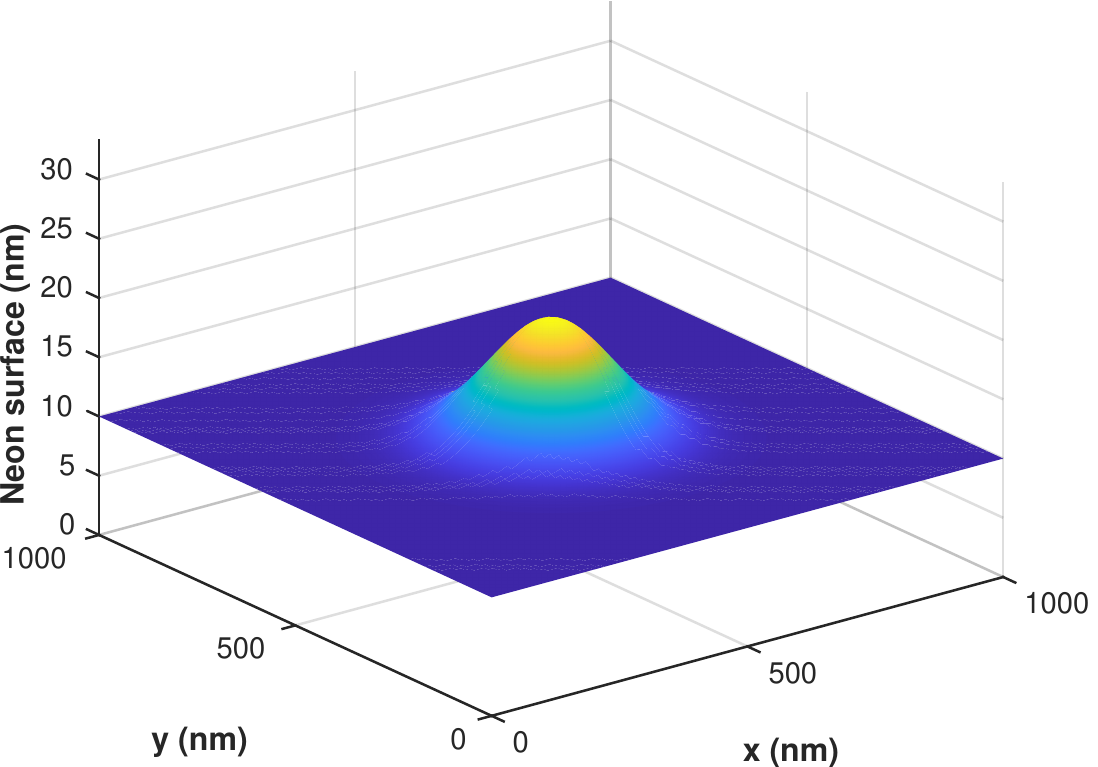}
        \caption{Bump surface}
        \label{fig:Fig1_a}
    \end{subfigure}\hfill
    \begin{subfigure}[t]{0.4\linewidth}
        \centering
        \includegraphics[width=\linewidth]{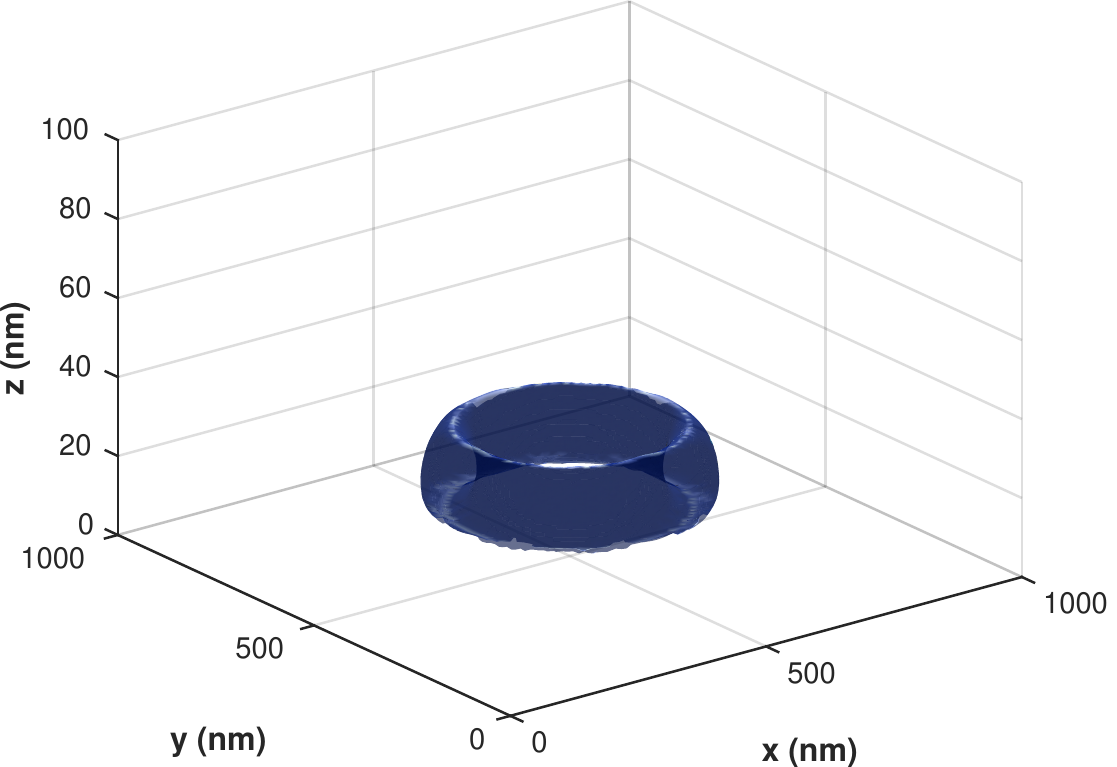}
        \caption{Bump wavefunction}
        \label{fig:Fig1_b}
    \end{subfigure}

    \vspace{0.6em}

    \begin{subfigure}[t]{0.4\linewidth}
        \centering
        \includegraphics[width=\linewidth]{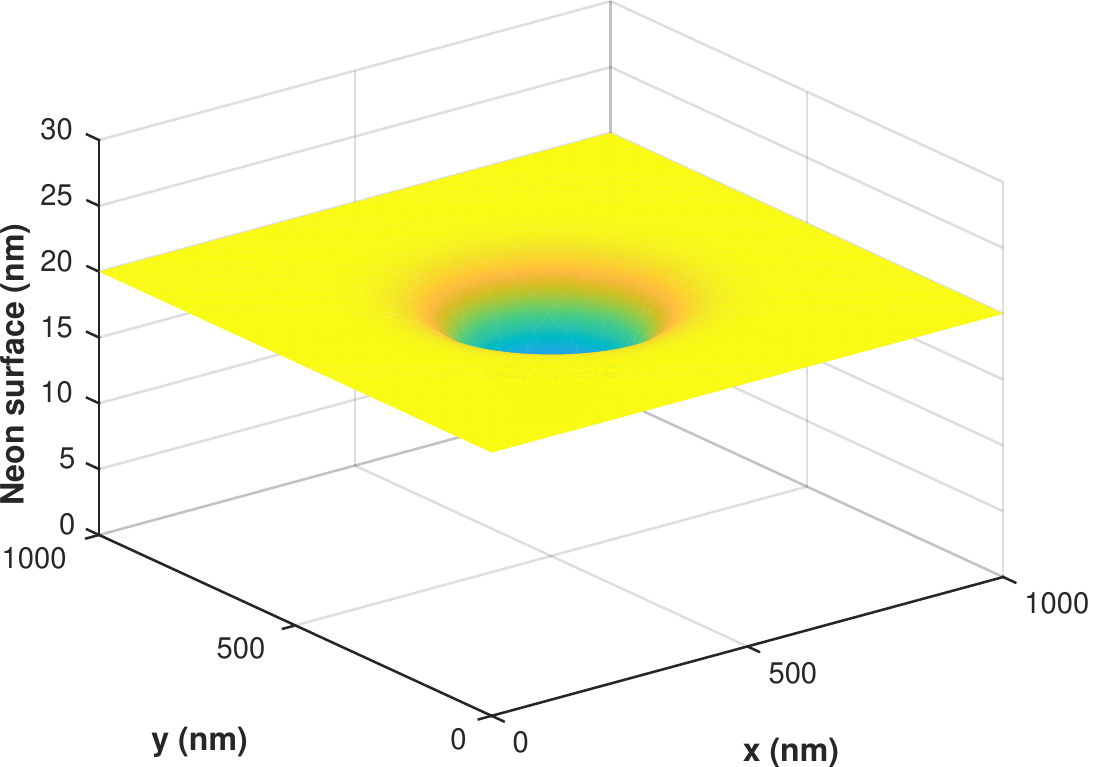}
        \caption{Valley surface}
        \label{fig:Fig1_c}
    \end{subfigure}\hfill
    \begin{subfigure}[t]{0.4\linewidth}
        \centering
        \includegraphics[width=\linewidth]{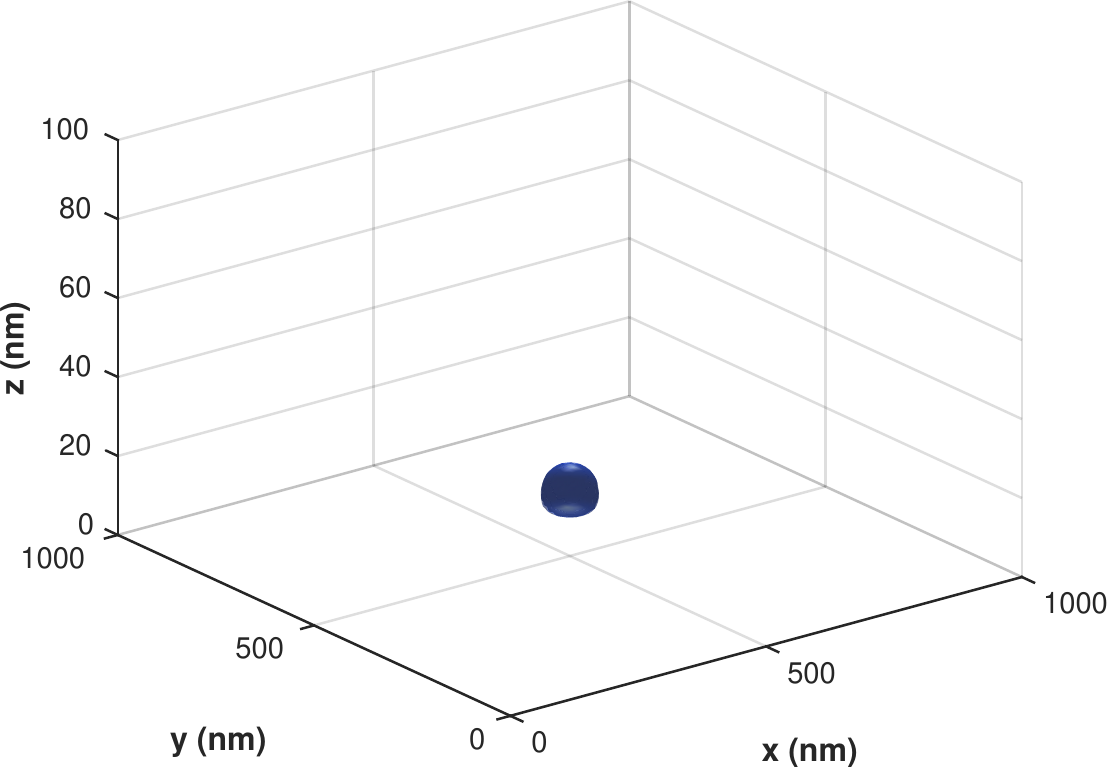}
        \caption{Valley wavefunction}
        \label{fig:Fig1_d}
    \end{subfigure}
    \caption{Representative bump and valley surface geometries and their corresponding electron wavefunctions. The bump produces a ring-like bound state around the protrusion, whereas the valley localizes the electron near the center of the depression.}
    \label{fig:Fig1}
\end{figure}

The simulated gaussian bump and valley geometries, together with their corresponding bound-state wavefunctions, are shown in Fig.~\ref{fig:Fig1}. For the bump geometry [Fig.~\ref{fig:Fig1}(a)], the electron wavefunction [Fig.~\ref{fig:Fig1}(b)] acquires a toroidal distribution surrounding the protrusion rather than peaking directly above its apex, consistent with the formation of an annular trapping potential. In contrast, the valley geometry [Fig.~\ref{fig:Fig1}(c)] produces a wavefunction [Fig.~\ref{fig:Fig1}(d)] that is localized near the center of the depression, indicating a qualitatively different confinement symmetry. These results confirm that both convex and concave surface perturbations can bind electrons, but with markedly different spatial character and spectral response.

\begin{figure}[t]
    \centering
    \begin{subfigure}[t]{0.32\linewidth}
        \centering
        \includegraphics[width=\linewidth]{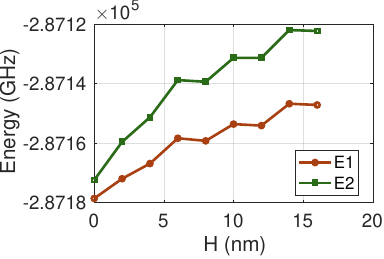}
        \caption{$E_1$, $E_2$ vs bump height $H$}
        \label{fig:B_E1E2_vs_H_both_absolute}
    \end{subfigure}\hfill
    \begin{subfigure}[t]{0.32\linewidth}
        \centering
        \includegraphics[width=\linewidth]{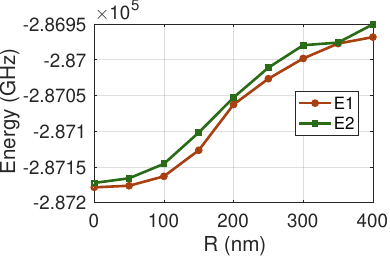}
        \caption{$E_1$, $E_2$ vs bump radius $R$}
        \label{fig:B_E1E2_vs_R_both_absolute}
    \end{subfigure}\hfill
    \begin{subfigure}[t]{0.32\linewidth}
        \centering
        \includegraphics[width=\linewidth]{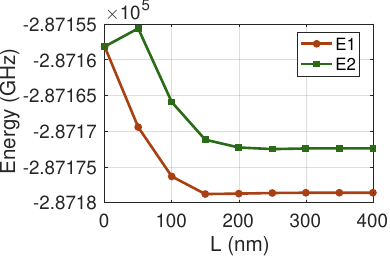}
        \caption{$E_1$, $E_2$ vs bump displacement $L$}
        \label{fig:B_E1E2_vs_L_both_absolute}
    \end{subfigure}
    \caption{Dependence of the two lowest orbital energies on bump geometry parameters: (a) height $H$, (b) radius $R$, and (c) displacement $L$ of bump from the gate center. Increasing $H$ and $R$ raises the orbital energies, whereas increasing $L$ reduces the bump-induced perturbation and lowers the energies toward saturation.}
    \label{fig:Bump_E1E2_subfigures}
\end{figure}

\begin{figure}[t]
    \centering
    \begin{subfigure}[t]{0.32\linewidth}
        \centering
        \includegraphics[width=\linewidth]{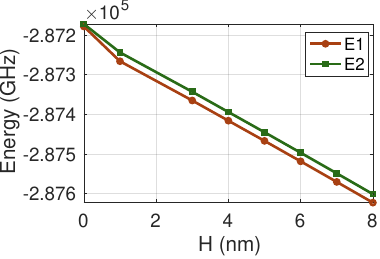}
        \caption{$E_1$, $E_2$ vs valley depth $H$}
        \label{fig:V_E1E2_vs_H_both_absolute}
    \end{subfigure}\hfill
    \begin{subfigure}[t]{0.32\linewidth}
        \centering
        \includegraphics[width=\linewidth]{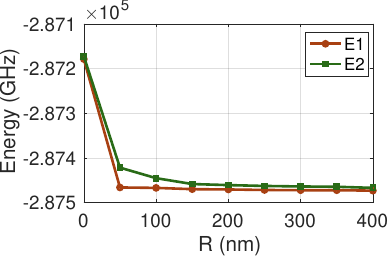}
        \caption{$E_1$, $E_2$ vs valley radius $R$}
        \label{fig:V_E1E2_vs_R_both_absolute}
    \end{subfigure}\hfill
    \begin{subfigure}[t]{0.32\linewidth}
        \centering
        \includegraphics[width=\linewidth]{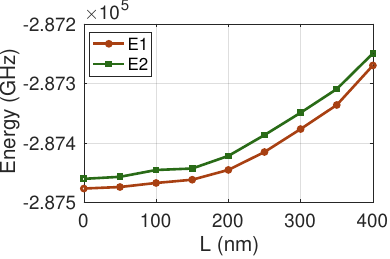}
        \caption{$E_1$, $E_2$ vs valley displacement $L$}
        \label{fig:V_E1E2_vs_L_both_absolute}
    \end{subfigure}
    \caption{Dependence of the two lowest orbital energies on valley geometry parameters: (a) depth $H$, (b) radius $R$, and (c) off-center displacement $L$. Increasing $H$ and $R$ lowers the orbital energies, whereas increasing $L$ reduces the trapping efficiency and raises the energies.}
    \label{fig:Valley_E1E2_subfigures}
\end{figure}

To quantify the sensitivity of the trap to surface roughness, each perturbation was parameterized by its amplitude $H$ (bump height or valley depth), lateral radius $R$, and off-center displacement $L$ from the trap gate center. For bumps, the $H$ and $R$ sweeps were performed at a fixed bump displacement $L=100\,\mathrm{nm}$, and increasing either parameter shifts both $E_1$ and $E_2$ upward, reflecting a weakening of the intended central confinement. In this geometry, the bump-induced state is more accurately described as a two-lobed, $p$-like orbital rather than an ideal ring state, owing to the anisotropic confinement of the elliptical gate together with the small circular Gaussian bump. The step-like features in Fig.~\ref{fig:B_E1E2_vs_H_both_absolute} and the apparent near-degeneracies in Fig.~\ref{fig:B_E1E2_vs_R_both_absolute} arise from discretization of the Gaussian surface on the finite simulation mesh, so adjacent nominal parameters can produce nearly identical effective topographies. In Fig.~\ref{fig:B_E1E2_vs_L_both_absolute}, the $L=0$ case is only nearly degenerate, with a very small residual splitting; displacing the bump breaks the symmetry, increases the level separation, and then drives both energies toward saturation as the displaced bump progressively modifies the local potential profile experienced by the electron. As shown in Fig.~\ref{fig:Valley_E1E2_subfigures} valleys show the opposite behaviour: increasing depth or radius lowers both orbital energies by strengthening confinement within the depression, whereas increasing $L$ shifts the valley-induced potential minimum away from the trap center, weakening the effective confinement profile experienced by the electron and thereby raising both orbital energies. Together, these results show that even modest nanoscale roughness can substantially reshape the trapping landscape and modify the low-energy orbital spectrum.

\section{Roughness-Induced Modification of the Quantum Confinement Landscape}

\begin{figure}[H]
    \centering
    \begin{subfigure}{0.95\linewidth}
        \centering
        \includegraphics[width=\linewidth]{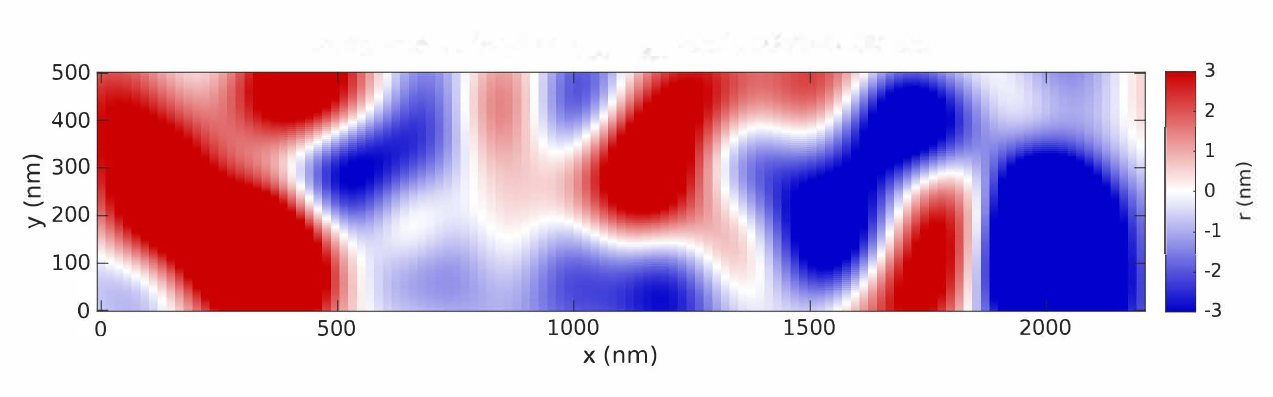}
        \caption{Random neon surface roughness profile with RMS height $\sigma_{\mathrm{rms}}=3~\mathrm{nm}$.}
        \label{fig:random_roughness_profile}
    \end{subfigure}

    \vspace{0.4cm}

    \begin{subfigure}{0.75\linewidth}
        \centering
        \includegraphics[width=\linewidth]{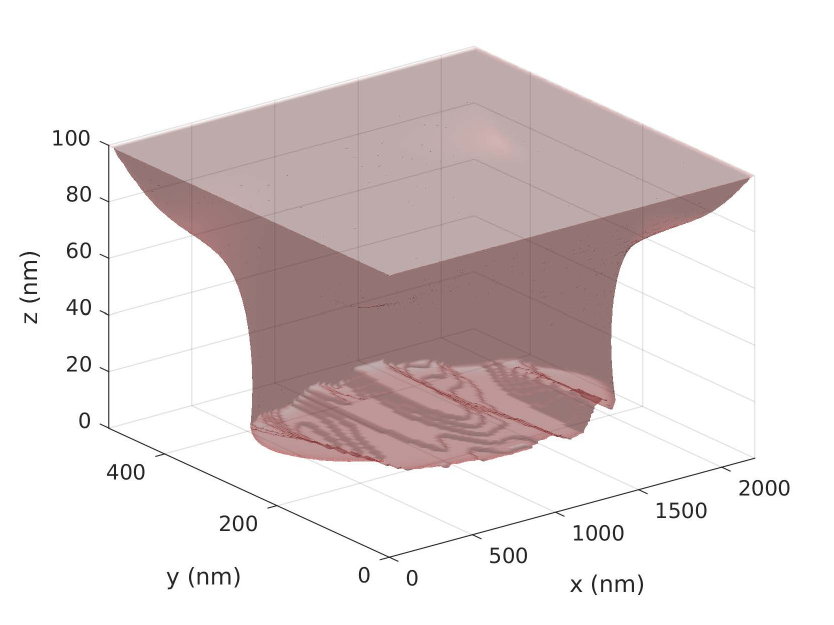}
        \caption{Three-dimensional isosurface of the resulting quantum potential $V_q$ for the rough neon surface.}
        \label{fig:random_roughness_Vq}
    \end{subfigure}

    \caption{Random neon surface roughness simulation and the corresponding three-dimensional confinement landscape.}
    \label{fig:random_neon_roughness}
\end{figure}

To extend the analysis beyond idealized bumps and valleys, we simulated an approximate neon surface-roughness profile motivated by the thin-film growth characteristics reported in Ref.~\cite{Matkovic2025NeonThinFilm}. However, it did not provide a direct nanoscale map of the surface topography, we therefore constructed a representative roughness profile with $\sigma_{\mathrm{rms}}=3~\mathrm{nm}$ and amplitudes clipped at $\pm 3~\mathrm{nm}$. As shown in Fig.~\ref{fig:random_roughness_profile}, the resulting nonuniform surface contains multiple shallow hills and depressions, which strongly modify the confinement landscape near the neon interface which can be seen in Fig.~\ref{fig:random_roughness_Vq}. For this realization, the orbital splitting increases from the smooth-surface value of $\Delta E \sim 6.2~\mathrm{GHz}$ to $\Delta E\approx10.6~\mathrm{GHz}$, indicating that plausible nanoscale roughness may shift the qubit transition frequency and produce unintended localization sites.

\section{Dielectric Interlayer for Roughness Suppression}

\begin{figure}[H]
    \centering

    \begin{subfigure}[b]{0.46\linewidth}
        \centering
        \includegraphics[width=1\linewidth]{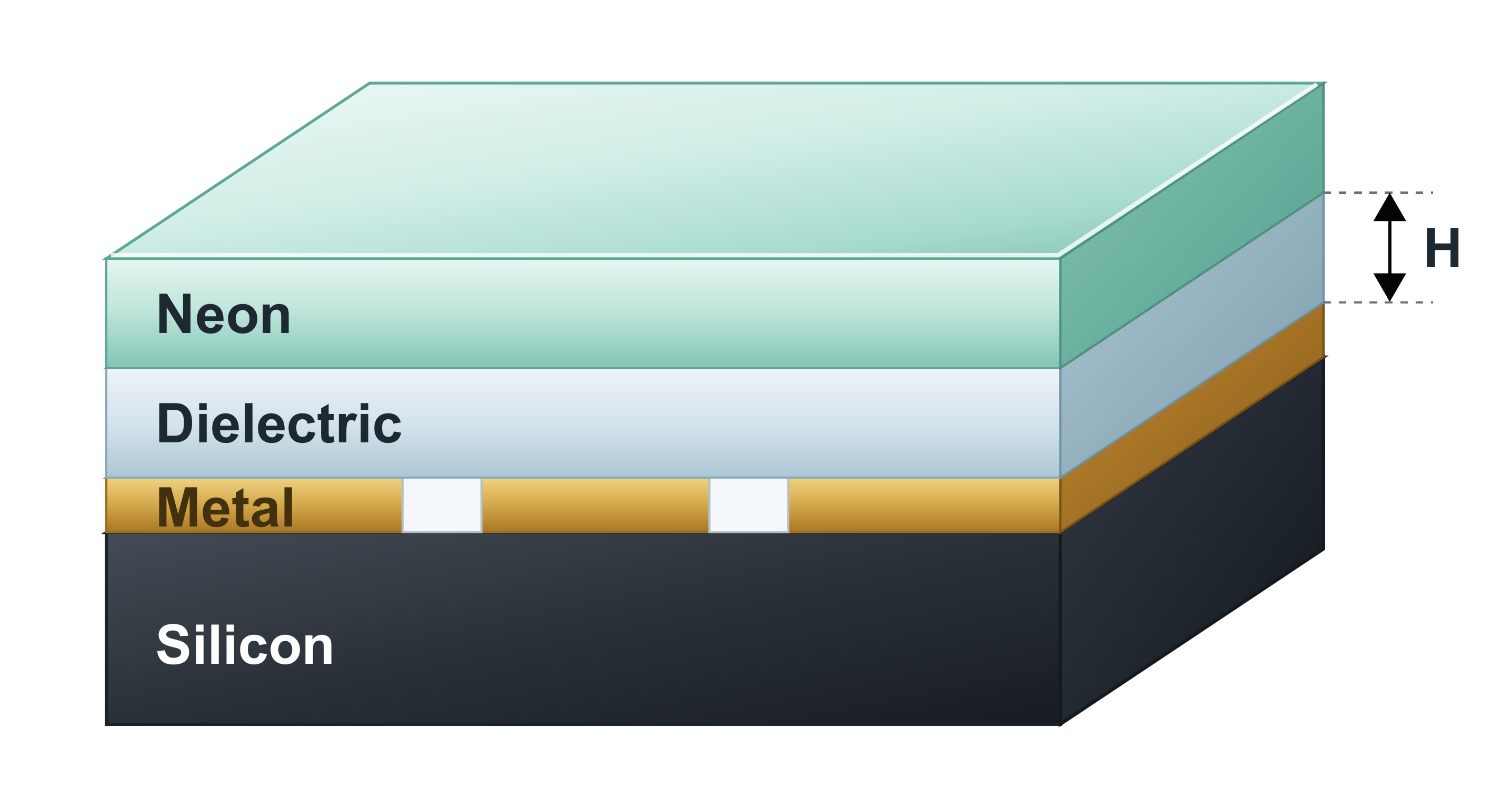}
        \caption{Device layer stack shown with neon deposited on dielectric layer. }
        \label{fig:hBN_device_geom}
    \end{subfigure}
    \hfill
    \begin{subfigure}[b]{0.46\linewidth}
        \centering
        \includegraphics[width=\linewidth]{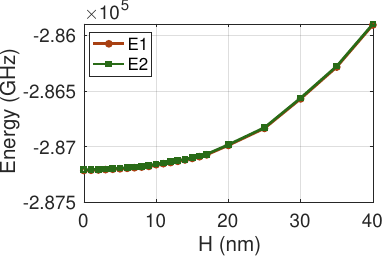}
        \caption{Dependence of the two lowest orbital energies $E_1$ and $E_2$ on dielectric thickness $H$ for $\epsilon_r = 3.5$.}
        \label{fig:hBN_E1E2_vs_H_sub}
    \end{subfigure}
    
    \vspace{0.5cm}
    \begin{subfigure}[b]{0.4\linewidth}
        \centering
        \includegraphics[width=\linewidth]{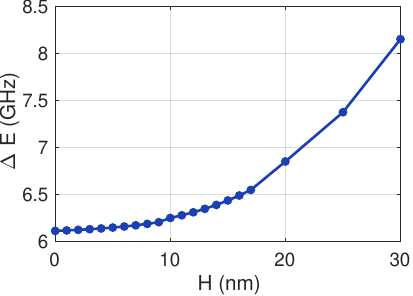}
        \caption{Dependence of qubit frequency as a function of dielectric thickness $H$ for $\epsilon_r = 3.5$ }
        \label{fig:DeltaE_vs_H}
    \end{subfigure}
    \hfill
    \begin{subfigure}[b]{0.4\linewidth}
        \centering
        \includegraphics[width=\linewidth]{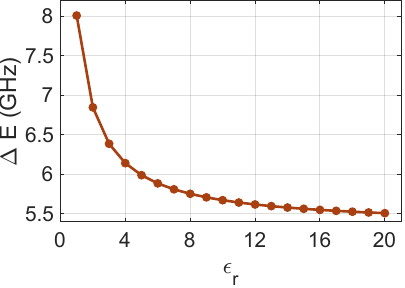}
        \caption{Dependence of qubit frequency as a function of dielectric constant $\epsilon_r$ }
        \label{fig:DeltaE_vs_epsilon_r}
    \end{subfigure}

    \caption{(a) Device stack with dielectric layer. (b) Variation of the two lowest orbital energies with dielectric thickness $H$. (c) $\Delta E$ vs $H$ (d) $\Delta E$ vs $\epsilon_r$.}
    \label{fig:hBN_device_and_E1E2}
\end{figure}

A major challenge in solid-neon electron devices is the unintended trapping of electrons by nanoscale surface roughness, such as bumps, valleys, and atomic-scale step features as analyzed in the previous section. These irregularities can locally distort the electrostatic landscape and generate parasitic bound states, leading to uncontrolled electron localization, charge noise, and additional decoherence channels. To mitigate this effect, a thin spacer layer of dielectric can be introduced between the patterned metallic gate structure and the neon film as shown in Fig. \ref{fig:hBN_device_geom}. In this geometry, the device consists of a buried metallic gate at the bottom, a dielectric spacer deposited above the gate to planarize the underlying topography, a solid neon layer deposited on top of dielectric, and vacuum above the neon surface where the electron is confined. The role of dielectric is therefore twofold: it can provide a flat template for neon growth, and it spatially separates the electron from unintended inter-electrode gap trap between adjacent electrodes and local surface roughness.

To quantify this effect, a thickness sweep of the dielectric spacer was performed while monitoring the two lowest orbital energies, $E_1$ and $E_2$. The results are shown in Fig.~\ref{fig:hBN_E1E2_vs_H_sub}. As the dielectric thickness increases, both $E_1$ and $E_2$ shift monotonically upward. This trend indicates that increasing the thickness of the spacer reduces the penetration of the gate electric field into the neon--vacuum confinement region, thereby weakening the overall confinement potential and making the bound-state energies less negative. We also analyzed  $\Delta E = E_2-E_1$ as a function of dielectric thickness H (Fig. \ref{fig:DeltaE_vs_H})and also how choice of dielectric $\epsilon_r$ affects qubit frequency (Fig. \ref{fig:DeltaE_vs_epsilon_r}). Physically, this reveals the central trade-off introduced by dielectric layer: a thicker spacer improves surface smoothness and suppresses roughness-induced trapping. The optimal choice of dielectric and its thickness must therefore balance surface polarization against sufficient field penetration for robust qubit confinement.

\begin{figure}[H]
    \centering
    \includegraphics[width=0.45\textwidth]{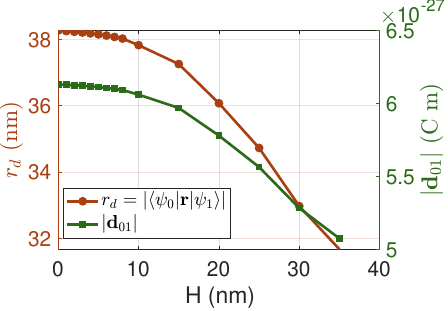}
    \caption{Transition dipole length $r_d$ (left axis) and electric dipole moment $|d_{01}|$ (right axis) as a function of dielectric thickness $H$.}
    \label{fig:hbn_dipole_vs_H}
\end{figure}

In addition to modifying the orbital energies, the dielectric spacer also affects the charge distribution relevant for charge--photon coupling. To quantify this effect, we calculated the transition dipole length between the ground and first excited orbital states using
\[
r_d = |\langle \psi_0|\mathbf{r}|\psi_1\rangle|,
\]
with the corresponding electric dipole moment given by $|d_{01}| = e *r_d$. As shown in Fig.~\ref{fig:hbn_dipole_vs_H}, both $r_d$ and $|d_{01}|$ decrease as the dielectric thickness increases. This behavior is consistent with the reduced penetration of the gate electric field through the thicker dielectric spacer, which weakens the electrostatic confinement and reduces the spatial overlap responsible for the transition dipole. 

\section{Electron Trapping Enabled by a Dielectric-Etched Trench}

\begin{figure}[H]
    \centering

    \begin{subfigure}[t]{0.45\linewidth}
        \centering
        \includegraphics[width=1\linewidth]{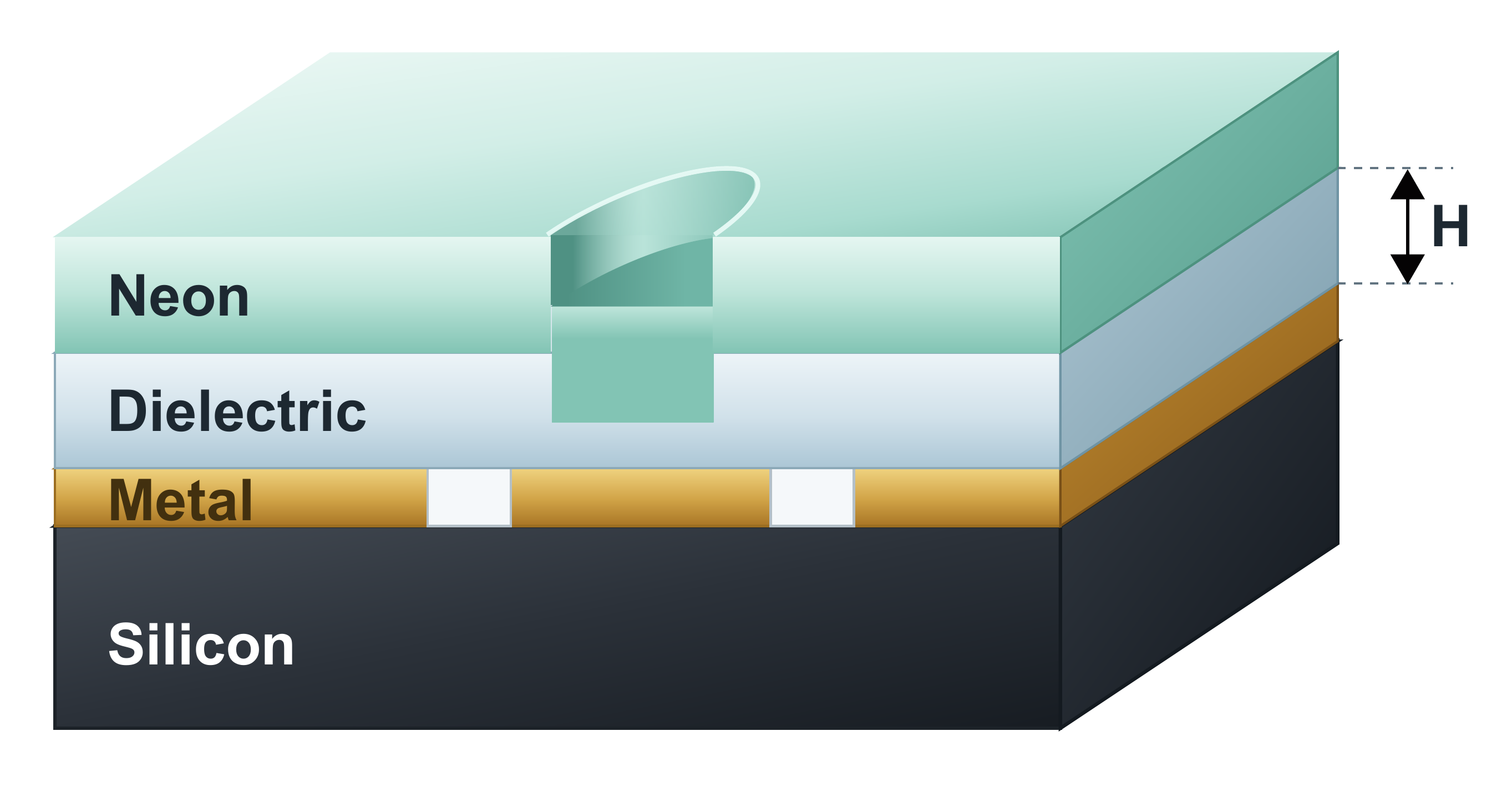}
        \caption{Device stack with dielectric etched quantum dot.}
        \label{fig:qd_device_pot}
    \end{subfigure}
    \hfill
    \begin{subfigure}[t]{0.45\linewidth}
        \centering
        \includegraphics[width=\linewidth]{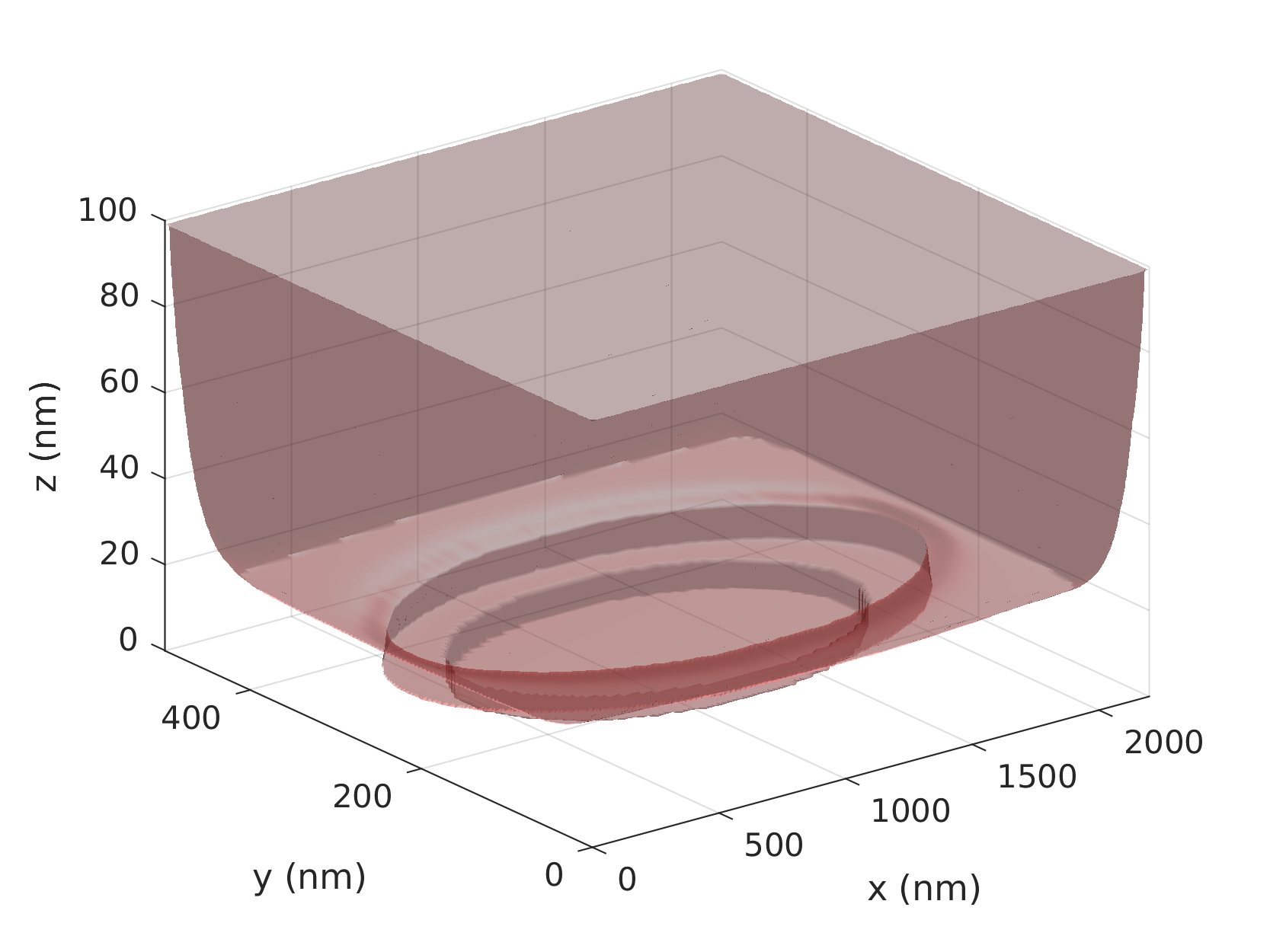}
        \caption{Three-dimensional isosurface of the confinement potential.}
        \label{fig:hBNTrenchPotential3D}
    \end{subfigure}

    \vspace{0.6em}

    \begin{subfigure}[t]{0.45\linewidth}
        \centering
        \includegraphics[width=\linewidth]{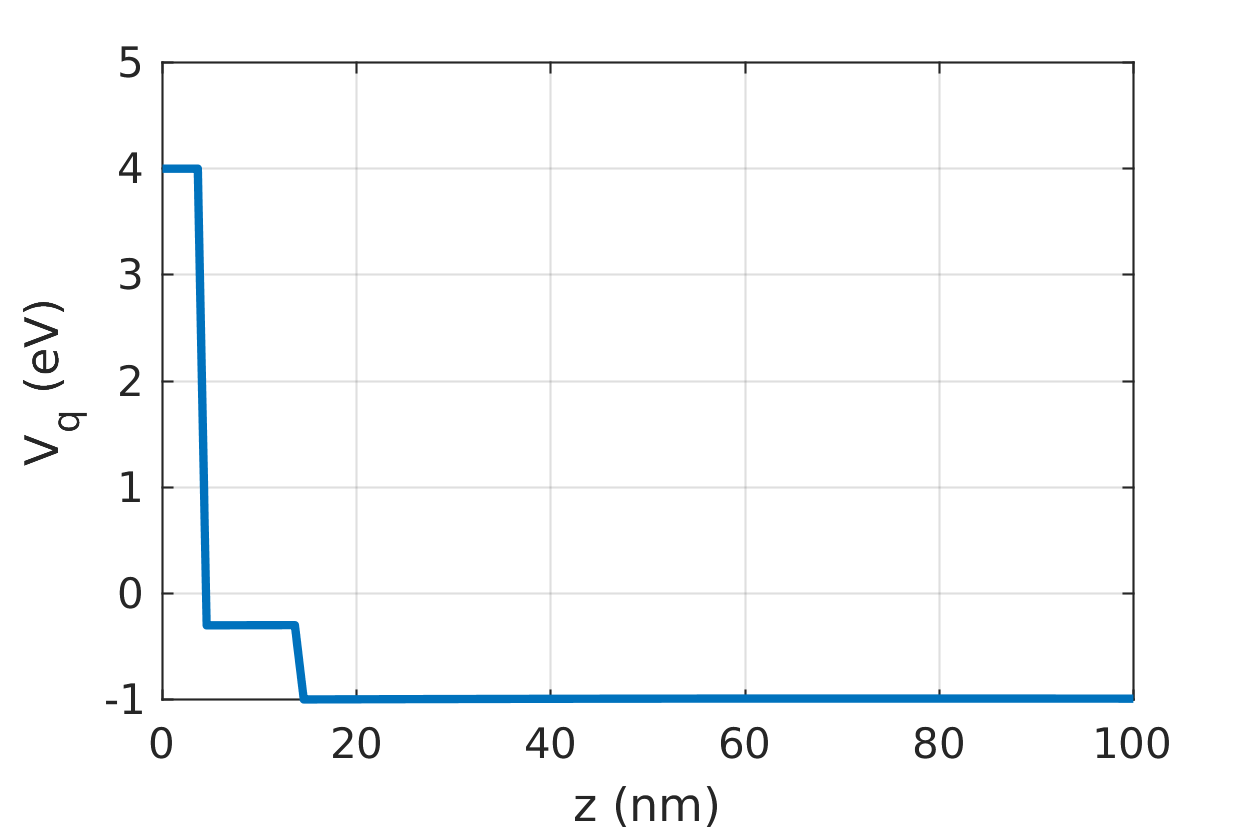}
        \caption{Potential profile along the central $z$-line.}
        \label{fig:hBNTrenchPotential1Dx}
    \end{subfigure}
    \hfill
    \begin{subfigure}[t]{0.45\linewidth}
        \centering
        \includegraphics[width=\linewidth]{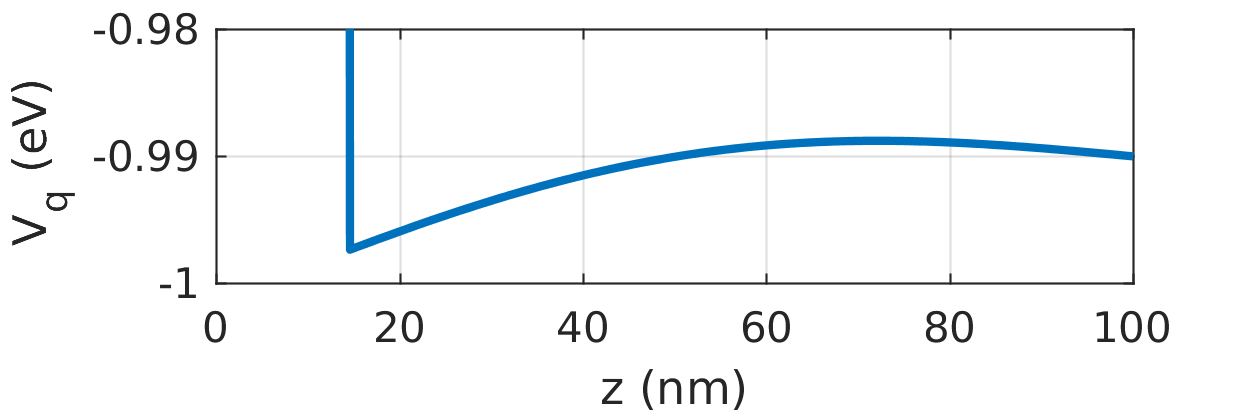}
        \caption{Zoomed potential profile.}
        \label{fig:hBNTrenchPotential1Dz}
    \end{subfigure}

    \caption{Device geometry and electrostatic confinement produced by the dielectric-etched trench structure. The etched dielectric layer  locally enhances gate-field penetration and creates a controlled trapping region in the neon--vacuum confinement landscape.}
    \label{fig:hBNTrenchPotential}
\end{figure}

While a continuous dielectric spacer helps suppress roughness-induced trapping by planarizing the surface, selective etching of dielectric can also be used as a deliberate tool for deterministic electron trapping mechanism. In this geometry, an elliptical metallic gate is buried underneath the dielectric stack, a dielectric layer of 10~nm is deposited above the gate, and an elliptical trench of 6~nm is etched into the dielectric before the neon film is deposited on top as shown in Fig. \ref{fig:qd_device_pot}. The neon layer then conforms to the etched profile, while vacuum above the neon provides the final electron confinement region. In the present design, the trench dimensions were chosen to be $0.75$ times the semi-axes of the underlying elliptical gate that is $a_{\mathrm{gate}} = 750~\mathrm{nm}$ and $b_{\mathrm{gate}} = 112.5~\mathrm{nm}$, so that the etched region remains aligned with the gate-defined potential while introducing an additional local modulation of the electrostatic landscape. This architecture combines the smoothing benefits of dielectric with an intentional topographic feature that can guide the electron into a well-defined trapping region.

The resulting electrostatic potential is shown in Fig.~\ref{fig:hBNTrenchPotential}. The three-dimensional isosurface plot [Fig.~\ref{fig:hBNTrenchPotential3D}] reveals that the etched trench creates a closed, low-potential region above the device, indicating that the local reduction in dielectric thickness enhances field penetration and produces a lateral trapping contour. The one-dimensional cut through the device center [Fig.~\ref{fig:hBNTrenchPotential1Dx}] shows a nearly flat and strongly confining interior region bounded by steep walls near the trench edges, while the zoomed vertical profile [Fig.~\ref{fig:hBNTrenchPotential1Dz}] confirms that the confinement minimum is located in the neon--vacuum region above the etched structure. Together, these results show that the trench acts as a controlled electrostatic template that restores strong local confinement without reintroducing uncontrolled surface roughness.

\begin{figure}[H]
    \centering
    \includegraphics[width=0.95\linewidth]{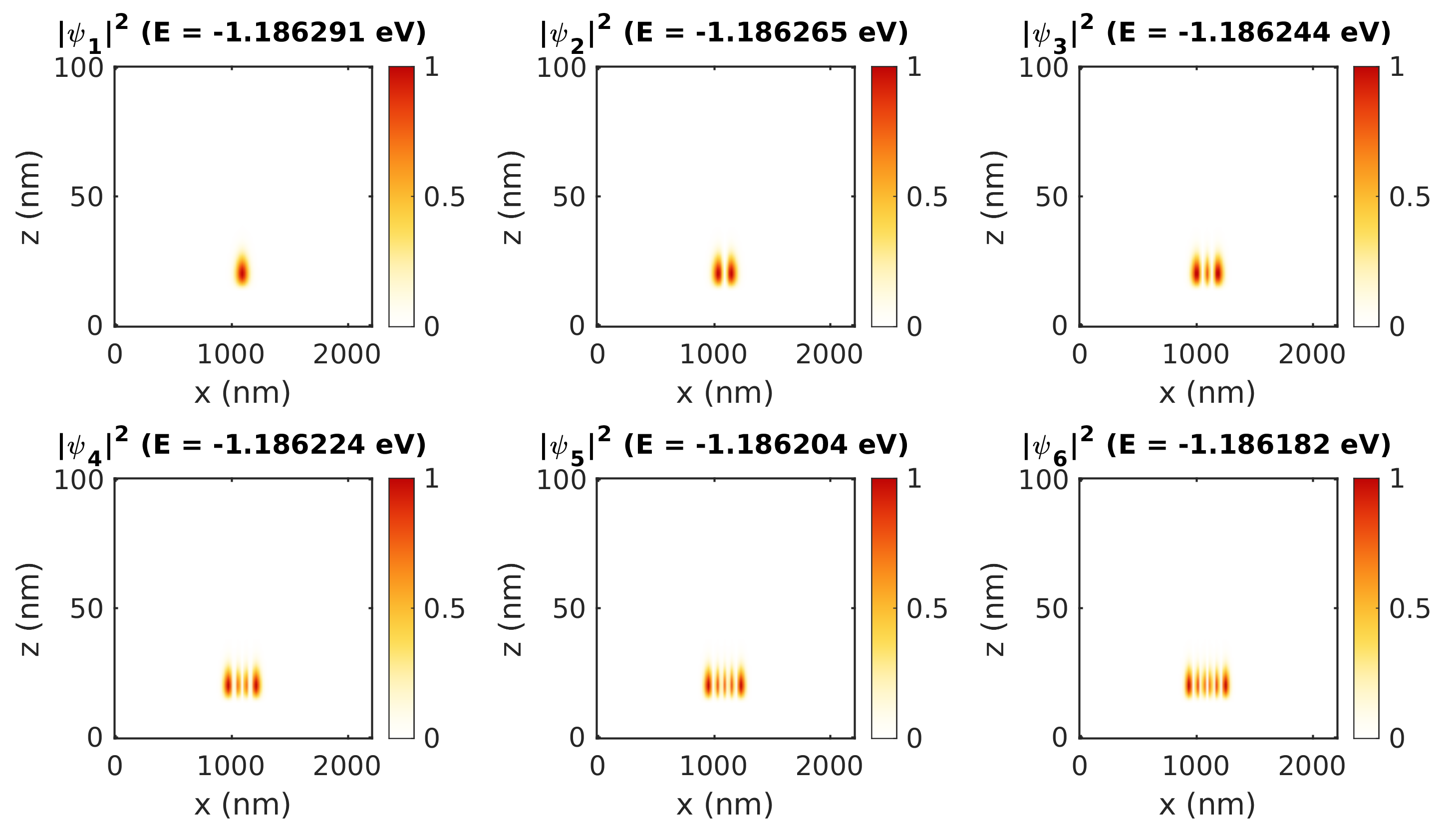}
    \caption{First six bound-state probability densities $|\psi_n|^2$ in the $x$--$z$ plane at mid-$y$ for the dielectric-etched trench geometry.}
    \label{fig:hBNTrenchWfs}
\end{figure}

\begin{table}[H]
    \centering
    \caption{First six eigenenergies for the dielectric-etched trench device, together with the splitting relative to the ground state.}
    \label{tab:hBN_trench_eigs}
    \begin{tabular}{c c c c}
        \hline
        State $n$ & $E_n$ (eV) & $E_n - E_1$ (meV) & $(E_n - E_1)$ (GHz) \\
        \hline
        1 & -1.187779425622 & 0.000000 & 0.0000 \\
        2 & -1.187754177283 & 0.025248 & 6.1050 \\
        3 & -1.187731339145 & 0.048086 & 11.6273 \\
        4 & -1.187708492459 & 0.070933 & 17.1516 \\
        5 & -1.187684903918 & 0.094522 & 22.8553 \\
        6 & -1.187660604519 & 0.118821 & 28.7308 \\
        \hline
    \end{tabular}
\end{table}
The first six bound-state wavefunctions are shown in Fig.~\ref{fig:hBNTrenchWfs}. The ground state remains strongly localized within the trench-defined confinement region, while the excited states exhibit a progressively increasing number of lateral nodes, consistent with a quantized sequence of orbital modes within the elongated trap. The corresponding eigenenergies are listed in Table~\ref{tab:hBN_trench_eigs}. Importantly, the orbital spacing remains of the same order as in the baseline smooth-gate device, showing that the etched-dielectric approach can recover a comparable $\Delta E$ while simultaneously suppressing parasitic trapping from random surface disorder. This indicates that selective dielectric etching offers a practical route to deterministic electron trapping: the continuous dielectric layer smooths the surface globally, and the etched trench reintroduces confinement only where it is intentionally required.

\section{Conclusion and Future Prospects}

In this work, we have shown that the surface geometry of the solid-neon platform can be elevated from a passive fabrication detail to an active quantum-engineering parameter. Using a three-dimensional Schr\"odinger--Poisson framework, we first validated the baseline gate-defined levitating-electron trap and then demonstrated that nanoscale surface perturbations such as bumps and valleys can strongly reshape the confinement landscape, creating unintended bound states with distinct spatial symmetries. We further showed that introducing a dielectric spacer provides an effective route to suppress roughness-induced trapping by planarizing the surface, while selective etching of dielectric reintroduces confinement only at intended locations. In this way, the combination of global smoothing and local topographic patterning enables deterministic electron trapping without sacrificing the low-energy orbital structure required for qubit operation.

A central outcome of this study is that dielectric spacer can reduce unintentional electron trapping thereby reducing decoherence mechanisms but also create deterministic electron traps through the engineered dielectric interface. More broadly, these results suggest that surface-defined confinement may offer a scalable pathway for positioning electrons reproducibly while preserving the exceptional material cleanliness that makes the solid-neon platform attractive for quantum information processing.

Looking ahead, experimental realization of dielectric-smoothed and selectively etched neon devices will be important for validating deterministic electron loading and confinement stability. Future work should extend the present model to include spin-dependent dynamics and experimentally relevant device designs.

\section*{Acknowledgements}

The computational work was performed using facilities provided by the National Computational Infrastructure (NCI), supported by the Australian Government, through the UNSW HPC Scheme (DOI: 10.26190/PMN5-7J50). Computing resources were also provided by NCI through the National Computational Merit Allocation Scheme (NCMAS) 2026 allocation.

During the preparation of this work, the authors used ChatGPT to improve the readability, language clarity, and formatting of the manuscript. The authors subsequently reviewed and edited the generated content as necessary and take full responsibility for the final content of the publication.

\bibliography{apssamp}

\end{document}